\documentclass[twocolumn,preprintnumbers,amsmath,amssymb,prl]{revtex4}
\usepackage{graphicx}
\usepackage{dcolumn}
\usepackage{bm}
\usepackage{amssymb}

\begin{document}

\title{Floquet control of dipolaritons in quantum wells}

\author{O. Kyriienko$^{1}$}
\author{O. V. Kibis$^{2}$}\email{Oleg.Kibis(c)nstu.ru}
\author{I. A. Shelykh$^{3,4}$}

\affiliation{$^1$The Niels Bohr Institute, University of
Copenhagen, Blegdamsvej 17, DK-2100 Copenhagen, Denmark}
\affiliation{$^2$Department of Applied and Theoretical Physics,
Novosibirsk State Technical University, Novosibirsk 630073,
Russia} \affiliation{$^3$Science Institute, University of Iceland
IS-107, Reykjavik, Iceland} \affiliation{$^4$ITMO University, St.
Petersburg 197101, Russia}

\begin{abstract}
We developed the theory of dipolaritons in semiconductor quantum
wells irradiated by an off-resonant electromagnetic wave (dressing
field). Solving the Floquet problem for the dressed dipolaritons,
we demonstrated that the field drastically modifies all
dipolaritonic properties. In particular, the dressing field
strongly effects on terahertz emission from the considered system.
The described effect paves the way for optical control of
prospective dipolariton-based terahertz devices.
\end{abstract}

\maketitle

\section{Introduction}
Advances in laser and microwave techniques made possible the use
of high-frequency fields as tools to control both atomic and
condensed-matter structures (so-called ``Floquet engineering''
based on the Floquet theory of periodically driven quantum
systems~\cite{Hanggi_98,Bukov2015,Blanes2009}). As a consequence,
properties of electrons strongly coupled to an electromagnetic
field --- also known as ``electrons dressed by the field''
(dressed electrons) --- are currently in the focus of attention of
the scientific community. Recently, the physical characteristics
of dressed electrons were analyzed for various nanostructures,
including quantum
wells~\cite{Wagner_10,Teich_13,Morina_15,Dini_16}, quantum
rings~\cite{Sigurdsson_14,Joibari_14,Koshelev_15},
graphene~\cite{Perez_14,Glazov_14,Kibis_16,Kristinsson_16}, etc.
Developing this scientific trend at the border of quantum optics
and physics of nanostructures, we present the theory of dressed
dipolaritons in semiconductor quantum wells (QWs).
\begin{figure}[!h]
\includegraphics[width=0.9\linewidth]{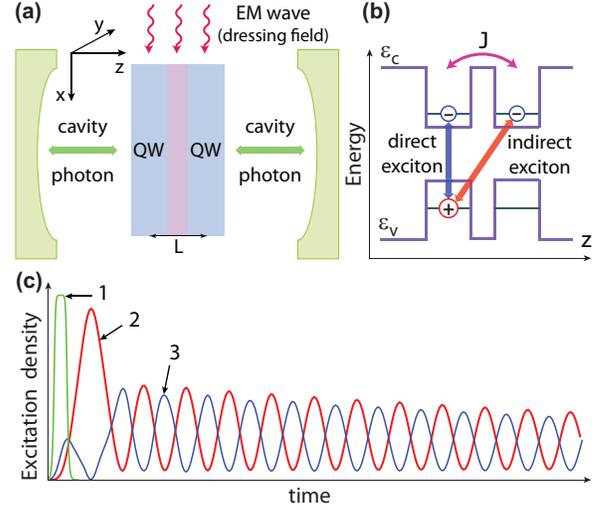}
\caption{Sketch of the system under consideration: (a) Two
tunnel-coupled semiconductor QWs with the inter-QW distance $L$,
which are embedded into a microcavity and irradiated by an
off-resonant electromagnetic wave (dressing field) incident along
the $x$ axis and linearly polarized along the $z$ axis; (b)
Structure of excited electron-hole pairs (direct and indirect
excitons) in QWs, where $\varepsilon_c$ is the conduction band,
$\varepsilon_v$ is the valence band, and $J$ denotes the tunnel
coupling of the QWs; (c) Density of excited cavity photons (curve
1), which results in oscillations of indirect and direct exciton
densities (curves 2 and 3, respectively).} \label{fig:sketch}
\end{figure}

Generally, polaritons are quasiparticles arisen from the strong
coupling between matter excitations and photons inside a
microcavity (see, e.g.,
Refs.~\cite{KavokinBook,CarusottoRev,DengRev}). They can be
controlled with different methods, including the electrical tuning
of the polarization \cite{Dreismann2016} and energy
\cite{Tsotsis2014}, the spatial tuning of the polariton condensate
by the excitonic reservoir engineering \cite{Askitopoulos2013},
the AC Stark tuning of the energy of polariton modes
\cite{Hayat2012,Cancellieri2014,Sie2015}, and others. If
tunnel-coupled semiconductor QWs are embedded into a microcavity
(see Fig.~1a), the strong coupling between photons in the cavity
and excitons in the QWs results in the formation of a particular
kind of polaritons known as dipolaritons~\cite{Cristofolini2012}.
The energy band structure hosting the dipolariton is pictured
schematically in Fig.~1b. The absorption of cavity photon excites
an electron-hole pair in a QW. In turn, the Coulomb attraction
between an electron and a hole forms a bound state (direct
exciton) confined in the same QW. Since there is the tunnel
coupling, $J$, between the two QWs, the conduction electron can
jump to the neighboring QW. The tunneling results in the
electron-hole pair confined in different QWs (indirect exciton)
which does not interact with cavity photons but couples to the
direct exciton (see, e.g.,
Refs.~\cite{High2012,ButovRev,Muljarov2012,Christmann2011}). Thus,
dipolariton is the three-component quasiparticle consisting from
the direct exciton, the indirect exciton and the cavity photon.

After the first experimental observation of
dipolaritons~\cite{Cristofolini2012}, it was demonstrated that
they have distinct response to electric and magnetic
fields~\cite{Wilkes2016} and stronger interparticle interaction as
compared to conventional polaritons~\cite{Byrnes2014}. Moreover,
they can be used for enhanced electrical
control~\cite{Rosenberg2016,Li2015}, facilitate indirect exciton
condensate preparation~\cite{Shahnazaryan2015}, single photon
emission~\cite{Kyriienko2014}, and other optoelectronic
applications. Particularly, it was predicted that dipolaritons can
serve as an efficient terahertz emission
source~\cite{Kyriienko2013,Kristinsson2013,Kristinsson2014}.
Namely, an excitation of the cavity photon mode by an optical
pulse (curve 1 in Fig.~1c) results in oscillations of indirect and
direct exciton densities (curves 2 and 3 in Fig.~1c, respectively)
with the THz frequency. It should be noted that indirect excitons
have large dipole moment along the $z$ axis, $d_z\approx eL$,
where $e$ is the electron charge and $L$ is the effective distance
between the centers of QWs (see Fig.~1a). Therefore, the
oscillations of the exciton density lead to the emission of
electromagnetic waves in the THz range.

Currently, the search for effective sources of THz emission is one
of the most exciting scientific trends at the border of applied
and fundamental physics. However, the use of dipolaritons as a
source of THz radiation needs effective methods to tune the THz
emission from dipolariton systems. In the present Letter, we
develop the theory of the optical control of dipolaritons with an
off-resonant laser excitation (dressing field), which creates
physical basis for such a tuning.

\section{Model}
Let us consider the conventional dipolariton setup consisting of
two tunnel-coupled semiconductor QWs embedded into a
microcavity~\cite{Cristofolini2012}. Additionally, we assume the
QWs to be irradiated by an off-resonant electromagnetic wave
(dressing field) incident along the $x$ axis and linearly
polarized along the $z$ axis (see Fig.~1a). It should be noted
that the cavity mirrors do not effect the propagation of the
dressing electromagnetic wave because of the chosen wave
orientation. The dressed dipolaritons in the QWs can be described
by the time-dependent Hamiltonian
\begin{eqnarray}\label{eq:H_0}
\hat{\cal H}(t)&=&\hbar\omega_1\hat{a}^\dagger_1\hat{a}_1+
\hbar\omega_2\hat{a}^\dagger_2\hat{a}_2+\hbar\omega_3\hat{a}^\dagger_3\hat{a}_3\nonumber
\\&+&\frac{\hbar\Omega_R}{2}(\hat{a}^\dagger_1\hat{a}_2+\hat{a}^\dagger_2\hat{a}_1)+\frac{J}{2}(\hat{a}^\dagger_2\hat{a}_3+\hat{a}^\dagger_3\hat{a}_2)\nonumber
\\&+&(d_zE\cos\omega t)\hat{a}^\dagger_3\hat{a}_3-
\frac{(d_{cv}E\cos\omega t)^2}{2 \hbar
\delta}\hat{a}^\dagger_2\hat{a}_2,
\end{eqnarray}
where $\hat{a}^\dagger_j$ and $\hat{a}_j$ are the creation and
annihilation operators, respectively, for a cavity photon ($j=1$),
a direct exciton ($j=2$), and an indirect exciton ($j=3$). The
physical meaning of the seven terms of the
Hamiltonian~(\ref{eq:H_0}) is as follows: The first, second and
third terms describe energies of noninteracting cavity photons,
direct excitons and indirect excitons with the frequencies
$\omega_{1,2,3}$, respectively; the fourth term describes the
interaction between a cavity photon and a direct exciton, which
results in the Rabi oscillations of this photon-exciton subsystem
with the Rabi frequency, $\Omega_R$; the fifth term describes the
tunnel coupling between direct and indirect excitons with the
coupling energy, $J$; the sixth term describes the usual dipole
interaction between an indirect exciton with the dipole moment,
$d_z$, and the electric field of the wave, $E\cos\omega t$, which
is actively studied in state-of-the-art experiments (see, e.g.,
Ref.~\cite{ButovRev}); the seventh term describes the
field-induced AC Stark shift of the direct
exciton~\cite{VonLehmen1986,Unold2004}. In contrast to the case of
indirect exciton, the dipole moment of the direct exciton is very
small. As a consequence, the dipole interaction between the direct
exciton and the wave is negligible weak. That is why the physical
origins of the sixth and seventh terms of the Hamiltonian
(\ref{eq:H_0}) are substantially different. It should be noted
also that the interband dipole moment, $d_{cv}$, is calculated
with using atomic wave functions. Therefore, the seventh term
almost does not depend on the orientation of the wave relative to
the growth direction of QW. As to the first five terms of the
Hamiltonian (\ref{eq:H_0}), they exactly coincide with the
conventional Hamiltonian describing ``bare''
dipolaritons~\cite{Cristofolini2012}.

To perform the Floquet analysis of the considered system, let us
apply the unitary transformation
$$\hat{U}(t)= \exp
\left[i \frac{\left(d_{cv}E\right)^2}{8\hbar^2 \omega \delta}
\sin2\omega t \,\,
\hat{a}^\dagger_2\hat{a}_2-i\frac{d_zE\sin\omega
t}{\hbar\omega}\hat{a}^\dagger_3\hat{a}_3\right]$$ to the
Hamiltonian (\ref{eq:H_0}). Then the transformed Hamiltonian of
dressed dipolaritons, $\hat{\cal H}^{'}(t)=
\hat{U}^\dagger(t)\hat{\cal H}(t)\hat{U}(t) -
i\hbar\hat{U}^\dagger(t)\partial_t \hat{U}(t)$, reads
\begin{align}\label{eq:H_1}
&\hat{\cal H}^\prime(t)=\hbar\omega_1\hat{a}^\dagger_1\hat{a}_1+
\left[\hbar\omega_2-\frac{(d_{cv} E)^2}{4\hbar
\delta}\right]\hat{a}^\dagger_2\hat{a}_2+\hbar\omega_3\hat{a}^\dagger_3\hat{a}_3\nonumber
\\&+\frac{J}{2}\cos\left[\frac{\left(d_{cv}E\right)^2}{8\hbar^2 \omega\delta}\sin2\omega
t +\frac{d_z E\sin\omega t}{\hbar\omega}\right]
(\hat{a}^\dagger_2\hat{a}_3+\hat{a}^\dagger_3\hat{a}_2)\nonumber
\\&+i\frac{J}{2}\sin\left[\frac{\left(d_{cv}E\right)^2}{8\hbar^2 \omega\delta}\sin2\omega
t+\frac{d_z E\sin\omega t}{\hbar\omega}\right]
(\hat{a}^\dagger_2\hat{a}_3-\hat{a}^\dagger_3\hat{a}_2)\nonumber
\\&+\frac{\hbar\Omega_R}{2}\cos\left[\frac{\left(d_{cv} E\right)^2}{8\hbar^2 \omega\delta}\sin2\omega
t
\right](\hat{a}^\dagger_1\hat{a}_2+\hat{a}^\dagger_2\hat{a}_1)\nonumber
\\&+i\frac{\hbar\Omega_R}{2}\sin\left[ \frac{\left(d_{cv}E\right)^2}{8\hbar^2 \omega\delta}\sin2\omega
t \right](\hat{a}^\dagger_1\hat{a}_2-\hat{a}^\dagger_2\hat{a}_1).
\end{align}
Within the conventional Floquet theory for periodically driven
quantum systems, the time-dependent Hamiltonian (\ref{eq:H_1}) can
be expanded into the power series of the inverse frequency,
$(1/\omega)^n$, where $n=0,1,2,...$ (the Floquet-Magnus
expansion~\cite{Bukov2015,Blanes2009}). Assuming the dressing
field to be high-frequency, $\omega\gg\Omega_R,J/\hbar$, one can
restrict the expansion to the zeroth-order term ($n=0$) which is
given by the Hamiltonian (\ref{eq:H_1}) averaged over the field
period, $T=2\pi/\omega$. As a result, we arrive from the
time-dependent Hamiltonian (\ref{eq:H_1}) at the effective
time-independent Hamiltonian of dressed dipolaritons, $\hat{\cal
H}_0=(1/T)\int_0^T\hat{\cal H}^\prime(t)dt$, which reads
\begin{eqnarray}\label{eq:H_2}
\hat{\cal H}_0&=&\hbar\omega_1\hat{a}^\dagger_1\hat{a}_1+
\hbar\widetilde{\omega}_2\hat{a}^\dagger_2\hat{a}_2+\hbar\omega_3\hat{a}^\dagger_3\hat{a}_3\nonumber
\\&+&\frac{\hbar\widetilde{\Omega}_R}{2}(\hat{a}^\dagger_1\hat{a}_2+\hat{a}^\dagger_2\hat{a}_1)
+\frac{\widetilde{J}}{2}(\hat{a}^\dagger_2\hat{a}_3+\hat{a}^\dagger_3\hat{a}_2),
\end{eqnarray}
where
\begin{align}
&\widetilde{\omega}_2=\omega_2- \frac{(d_{cv} E)^2}{4\hbar^2
\delta} ,\label{om}\\
&\widetilde{J}=\frac{J}{T}\int_0^T \exp \left[i
\frac{\left(d_{cv}E\right)^2}{8\hbar^2 \omega\delta} \sin2\omega
t + i\frac{d_zE\sin\omega t}{\hbar\omega}\right]dt,\label{J}\\
&\widetilde{\Omega}_R=\frac{\Omega_R}{T}\int_0^T \exp \left[i
\frac{\left(d_{cv}E\right)^2}{8\hbar^2 \omega\delta} \sin2\omega t
\right]dt.\label{Om}
\end{align}
%
%
\begin{figure}[h!]
\includegraphics[width=1.0\linewidth]{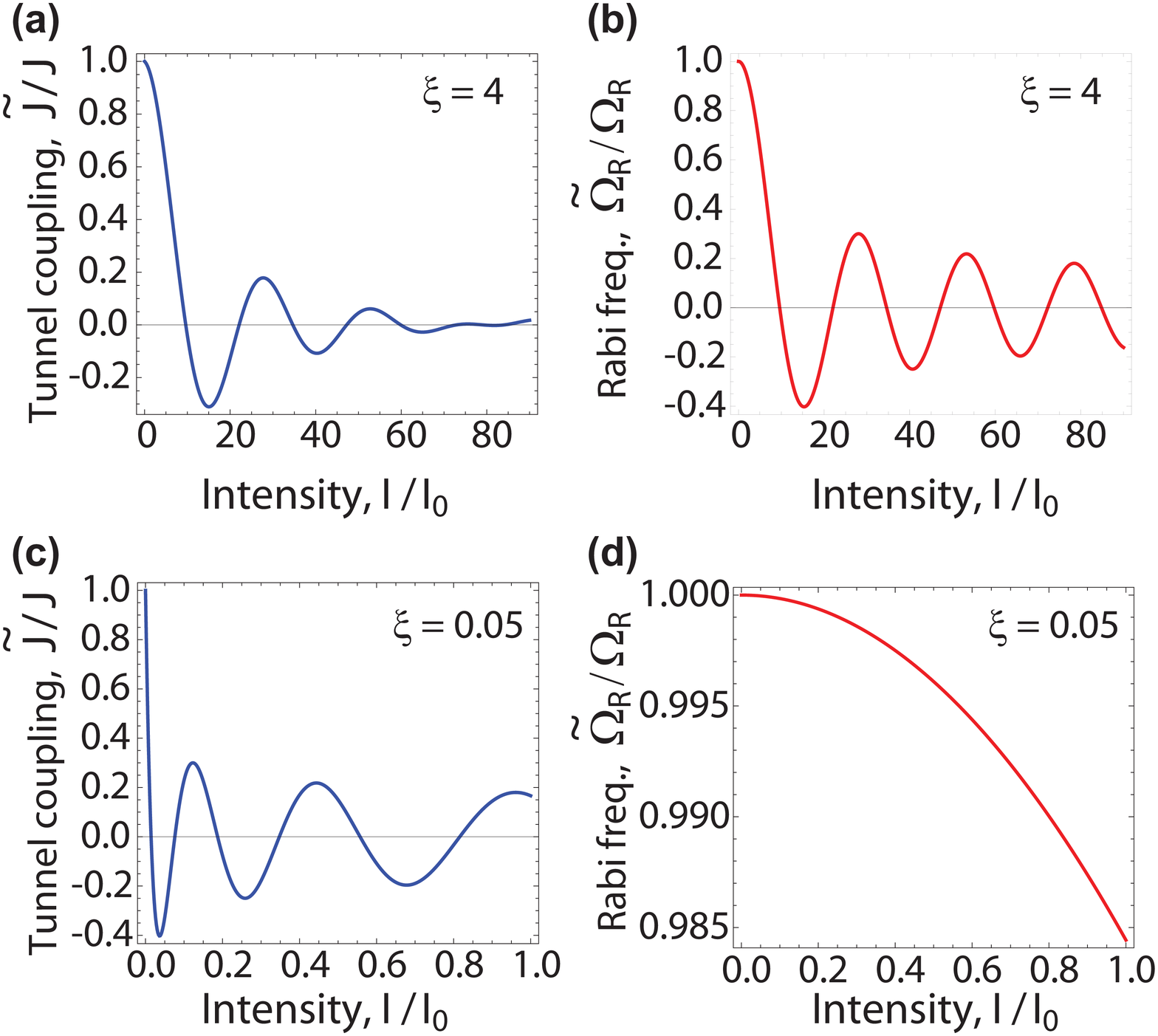}
\caption{Dependencies of the renormalized tunnel coupling,
$\widetilde{J}$, and Rabi frequency, $\widetilde{\Omega}_R$, on
the irradiation intensity, $I$, for different values of the
dipolariton-field coupling parameter, $\xi$. The intensity is
plotted in the units of $I_0=\hbar^2 |\delta| \omega c
\epsilon_0/d_{cv}^2$.} \label{fig:coupling}
\end{figure}
The Hamiltonian of dressed dipolaritons (\ref{eq:H_2}) exactly
coincides with the stationary Hamiltonian of ``bare'' dipolaritons
described by the first five terms of the Hamiltonian
(\ref{eq:H_0}) with the replacements
$\omega_2\rightarrow\widetilde{\omega}_2$,
$\Omega_R\rightarrow\widetilde{\Omega}_R$ and
$J\rightarrow\widetilde{J}$. Therefore, the quantities
(\ref{om})--(\ref{Om}) should be treated as stationary dipolariton
parameters renormalized by the dressing field.

\section{Discussion and conclusions}
First of all, let us discuss the dependence of the parameters
(\ref{om})--(\ref{Om}) on the dressing field amplitude, $E$, and
the frequency, $\omega$, restricting the consideration to the case
of the red-detuned dressing field ($\delta<0$). The field-induced
renormalization of the direct exciton  frequency (\ref{om}) is
described by the known dynamic Stark shift~\cite{VonLehmen1986}.
As to the renormalized parameters $\widetilde{J}$ and
$\widetilde{\Omega}_R$, the exponential functions in
Eqs.~(\ref{J})---(\ref{Om}) contain the two different terms which
are squared and linear in the field amplitude, $E$. Physically,
the first of them arises from the dressing of direct excitons,
whereas the second one is caused by the dressing of indirect
excitons. The coupling parameter of the first dressing depends on
the interband dipole moment, $d_{cv}$, and the detuning, $\delta$,
whereas the second one is described by the inter-QW dipole moment,
$d_z$. Therefore, it is reasonable to introduce the dimensionless
dipolariton-field coupling parameter,
$\xi=(d_{cv}/d_z)\sqrt{\omega/2|\delta|}$, which describes the
relative contribution of these two different dressing mechanisms
to the renormalization of dipolaritonic properties. This parameter
can be controlled by varying the inter-QW distance, $L$, and the
detuning, $\delta$. Particularly, the state-of-the-art
semiconductor technologies can easily fabricate QWs with the broad
range of the dipole moments, $0<d_{cv}/d_z<10$. Dependencies of
the renormalized tunnel coupling, $\widetilde{J}$, and the
renormalized Rabi frequency, $\widetilde{\Omega}_R$, on the
irradiation intensity, $I$, are plotted in Fig.~2 for different
values of the parameter $\xi$. We see that both tunnel coupling
and the Rabi frequency oscillate as functions of the irradiation
intensity. Mathematically, these oscillations arise from
periodical functions in Eqs.~(\ref{J})--(\ref{Om}). It follows
from Eq.~(\ref{Om}) that the renormalization of the Rabi
frequency, $\widetilde{\Omega}_R$, is described by the interband
dipole moment, $d_{cv}$, which does not depend on the sample size,
$L$. Therefore, the Rabi frequency, $\widetilde{\Omega}_R$, does
not depend on the parameter $\xi$ for a given semiconductor
material (see Figs.~2b and 2d). On the contrary, the renormalized
tunnel coupling (\ref{J}) depends on both the interband dipole
moment, $d_{cv}$, and the inter-QW dipole moment, $d_z$. Applying
the well-known Jacobi-Anger expansion to transform the exponential
function in Eq.~(\ref{J}), we arrive at the simple expression,
$\widetilde{J}\approx J\mathcal{J}_0 \left({d_z
E}/{\hbar\omega}\right)$, describing the tunnel coupling for
$\xi\ll1$,  where $\mathcal{J}_0$ is the zeroth-order Bessel
function of the first kind. It follows from this that the decrease
of $d_z$ leads to decreasing the period of oscillations of the
tunnel coupling, $\widetilde{J}$, as a function of the field
intensity, $I$ (see Fig.~2a and 2c). One can conclude that the
dressing field changes the intrinsic parameters of the
dipolariton, $J$ and $\Omega_R$, whereas the alternative approach
to control polaritonic systems by the interaction with reservoir
excitons results in changing dipolariton dynamics (see, e.g.,
Ref.~\cite{Askitopoulos2013}). Thus, these two methods of
dipolariton control supplement each other and can be combined in
experiments.
\begin{figure}[!h]
\includegraphics[width=0.9\linewidth]{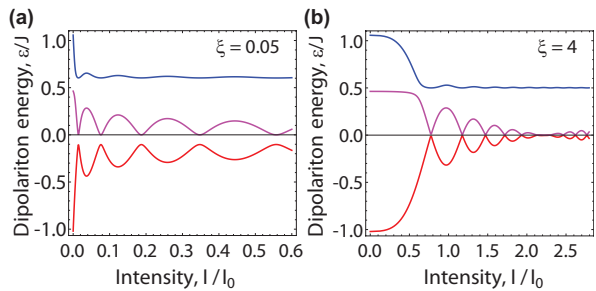}
\caption{The energy spectrum of dressed dipolariton,
$\varepsilon$, as a function of the dressing field intensity, $I$,
for $\omega_1=0.5J$, $\widetilde{\omega}_2=\omega_3=0$,
$\Omega_R=0.25J$.} \label{fig:spectrum}
\end{figure}

The discussed renormalization of the parameters
(\ref{om})--(\ref{Om}) by a dressing field results in the
renormalization of all physical properties of the dipolariton
system. Diagonalizing the Hamiltonian (\ref{eq:H_1}), we can
calculate numerically the energy spectrum of dressed dipolaritons,
which is shown in Fig.~3. We observe that the energy spectrum
consists of three energy levels which depend differently on the
dressing field intensity, $I$. Particularly, the dressing field
decreases the gap between the two lowest dipolariton energies (see
Fig.~3a) and can turn the gap into zero (see Fig.~3b). In order to
calculate the spectrum of electromagnetic emission from the
dipolariton system, we have to consider the dynamics of the system
under the pulsed excitation of the cavity mode (see curve 1 in
Fig.~1c) with an external source at the frequency $\omega_0$.
Introducing the detuning parameters for the photon-exciton modes
of the cavity, $\Delta_{1,2,3}=\hbar(\omega_{1,2,3}-\omega_0)$, we
can perform the Fourier analysis of the oscillations of exciton
densities within the approach used in
Refs.~\cite{Kyriienko2013,Kristinsson2014}. For definiteness, let
us focus on the dipolariton spectra of GaAs-based QWs with the
interband dipole moment $d_{cv}=28$~Debay and the exciton energy
$\hbar \omega_2 = 1.54$~eV, which are dressed by the field with
the frequency $\omega/2\pi=10^{13}$~Hz. As a result, one can
calculate numerically the spectrum of electromagnetic emission
from the excited dipolariton system, which is plotted in Fig.~4.
The spectrum consists of the main spectral line and the two
sideband spectral lines (see Fig.~4a). For typical samples, the
tunnel coupling, $J$, is of meV scale~\cite{Cristofolini2012}.
Thus, the main spectral line corresponds to the terahertz emission
(see Fig.~4b). It should be noted that one can construct a
single-mode terahertz emitter if the sideband modes are suppressed
with an additional THz cavity \cite{Kristinsson2014}. In the most
relevant case of $\xi\ll1$, the photon energies corresponding to
the main line and the sideband lines can be described adequately
by the expressions $\hbar\Omega\approx\widetilde{J}$ and
$\hbar\Omega\approx\Delta_1\pm\widetilde{J}$, respectively.
Therefore, we can write the frequency of terahertz emission
corresponding to the main spectral line, $\Omega$, as $
\hbar\Omega\approx J\mathcal{J}_0
\left({d_zE}/{\hbar\omega}\right). $ This means that the terahertz
emission from the dipolariton system is controlled by the dressing
field amplitude, $E$, and the dressing field frequency, $\omega$.
Particularly, varying the irradiation intensity, $I$, one can tune
the terahertz frequency, $\Omega$ (see Fig.~4b).
\begin{figure}
\includegraphics[width=0.9\linewidth]{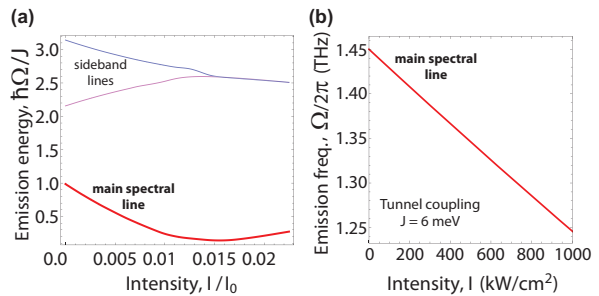}
\caption{Spectrum of electromagnetic emission, $\Omega$, from a
dipolariton system in GaAs-based QWs with the detuning
$\Delta_1=2.5J$, $\Delta_2=\Delta_3=0$, and the Rabi frequency
$\Omega_R=J$, as a function of irradiation intensity, $I$: (a)
Photon energy of the emission corresponding to the main and
sideband spectral lines; (b) Frequency of the terahertz emission.}
\label{fig:spectrum}
\end{figure}
%


Summarizing the aforesaid, one can conclude that an off-resonant
high-frequency electromagnetic field (dressing field) can be used
as an effective tool to control all physical properties of
dipolaritons in semiconductor QWs, including their energy spectrum
and dynamics. Particularly, the field-induced renormalization of
the coupling of direct and indirect excitons results in changing
terahertz emission from the dipolariton system. Since
light-controlled electronic devices are typically much faster than
those of electrically controlled, the proposed optical control of
dipolaritonic terahertz emitters is expected to be faster than the
usual gate control. Thus, the elaborated theory provides the
ground for novel optoelectronic devices operated by light.

\section{Funding}
RISE Program (CoExAN); FP7 ITN Program (NOTEDEV); Russian
Foundation for Basic Research (17-02-00053); Rannis (163082-051);
Ministry of Education and Science of Russian Federation
(3.4573.2017/6.7, 3.2614.2017/4.6, 14.Y26.31.0015); ERC QIOS
(306576).



\end{document}